\documentclass[a4paper,11pt]{article}

\usepackage[utf8]{inputenc}
\usepackage[T1]{fontenc}
\usepackage{lmodern}
\usepackage{geometry}
\usepackage{graphicx}
\usepackage{authblk}
\usepackage{url}
\usepackage{cite}
\usepackage{fancyhdr}
\usepackage{hyperref}
\usepackage{orcidlink}
\usepackage{amsmath,amssymb}
\usepackage{units}

\fancypagestyle{firstpage}{
	\fancyhf{}
	\fancyhead[C]{\includegraphics[height=2.2cm]{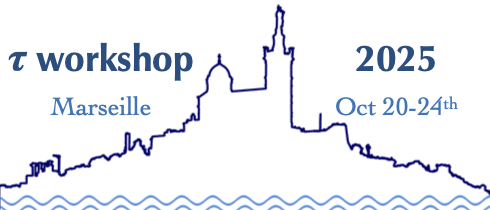}\\ \textit{Proceedings of the 18th International Workshop on Tau Lepton Physics}
	} 

}

\title{New avenues for tau flavor violation}
\author{Julian Heeck\,\orcidlink{0000-0003-2653-5962}}

\affil{Department of Physics, University of Virginia,
Charlottesville, Virginia 22904, USA}

\date{}

\begin{document}

\maketitle
\thispagestyle{firstpage}


\begin{abstract}
I present a concise overview of tau flavor violation and the broad opportunities it offers, with a focus on non-standard decay channels and their underlying theoretical motivations.
\end{abstract}

\section{Introduction}

\begin{figure}
\begin{center}
\includegraphics[width=0.9\textwidth]{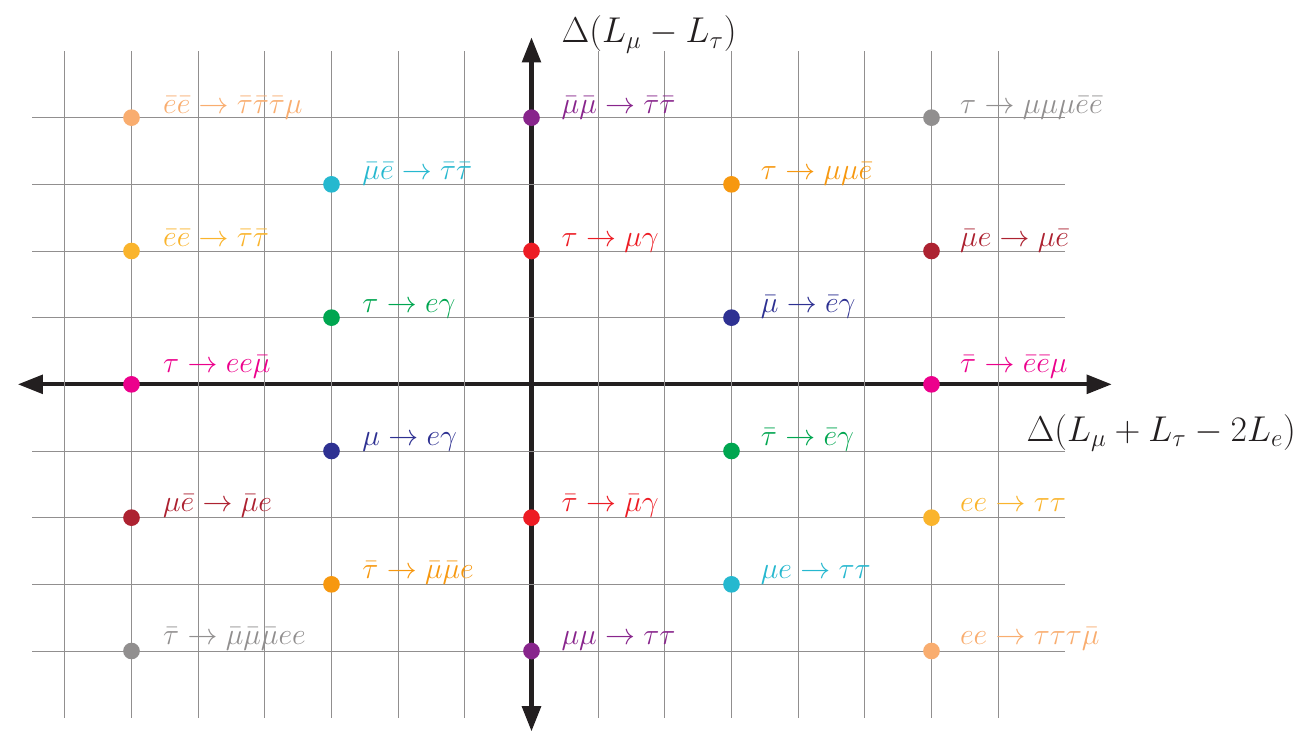}
\end{center}
\caption{
CLFV processes (only one representative shown per group) organized by their $U(1)_{L_\mu-L_\tau}\times U(1)_{L_\mu+ L_\tau -2 L_e}$ breaking, adapted from Ref.~\cite{Heeck:2016xwg}.
}
\label{fig:LFVgrid}
\end{figure}

Charged lepton flavor violation (CLFV) is forbidden in the Standard Model (SM) but allowed in many of its extensions, rendering it a useful way to search for new physics~\cite{Davidson:2022jai,Asadi:2025dii}. Mathematically, lepton flavor conservation is ensured by the two $U(1)$ symmetries $U(1)_{L_\mu-L_\tau}\times U(1)_{L_\mu + L_\tau - 2 L_e}$ of the SM Lagrangian~\cite{Heeck:2016xwg}. (Violations of the other two conserved SM quantities, baryon number~$B$ and total lepton number $L$, are not discussed here, even though they typically violate lepton flavor as well, see Ref.~\cite{Heeck:2024jei} for a recent study of tau flavor violation together with baryon number.)
CLFV processes can then be organized according to their $U(1)_{L_\mu-L_\tau}\times U(1)_{L_\mu + L_\tau - 2 L_e}$ breaking~\cite{Lew:1994zc}, illustrated in Fig.~\ref{fig:LFVgrid}. Here, each possible breaking pattern is shown with one representative CLFV process, e.g.~$\mu\to e\gamma$, even though there are infinitely many processes with the same CLFV pattern, e.g.~$\mu\to e\gamma$ comes together with $\mu\to 3e$, $\mu\to e \gamma\gamma$, $Z\to e\mu$, etc. \textit{All} of these processes are allowed if \textit{one} of them is allowed, with branching ratios determined by the underlying model. Which CLFV process to focus on to test a given CLFV group then depends on which model one wants to test, and they usually provide complementary information.

The two-dimensional nature of Fig.~\ref{fig:LFVgrid} immediately illustrates a very important fact: CLFV is not a simple yes or no question. To pin down \textit{how} lepton flavor is broken in our universe, we have to observe CLFV processes from \textit{at least} two different groups~\cite{Heeck:2016xwg}. Only observing, for example, $\mu\to e\gamma$ would not tell us if the orthogonal CLFV group $U(1)_{L_e + L_\mu - 2 L_\tau}$ is broken or not. In particular, we \textit{have to} observe CLFV involving tau leptons to fully understand lepton flavor!

The different CLFV breaking patterns in Fig.~\ref{fig:LFVgrid} cannot be meaningfully compared to each other in a model independent way, so searches for e.g.~$\mu\to e\gamma$ are inherently complementary to searches for $\tau\to \mu\gamma$ or $\tau\to \mu\mu\mu \bar{e}\bar{e}$. It is straightforward to build SM extensions with additional symmetries --  usually subgroups of $U(1)_{L_\mu-L_\tau}\times U(1)_{L_\mu + L_\tau - 2 L_e}$ -- in which many or even all but one group in Fig.~\ref{fig:LFVgrid} are forbidden, see Refs.~\cite{Heeck:2014qea,Heeck:2016xwg,Bigaran:2022giz,Heeck:2024uiz,Heeck:2025jfs}. It is then crucial to experimentally \textit{probe each group} in Fig.~\ref{fig:LFVgrid} to make sure we're not missing new physics due to theoretical bias towards the simplest breaking patterns, which make up the ``first ring'' around the origin in Fig.~\ref{fig:LFVgrid} and allow for simple experimentally-friendly two-body decays $\ell_\alpha\to \ell_\beta\gamma$, the focus of most CLFV studies. In the SM effective field theory (SMEFT), CLFV operators with this flavor content arise at mass dimension $d=6$, leading to amplitudes that are suppressed by two powers of the heavy mass scale: $\mathcal{M}\propto 1/\Lambda^2$. Current experiments are sensitive to $\Lambda \lesssim \mathcal{O}(\unit[10^3]{TeV})$ in the muon sector and $\Lambda \lesssim \mathcal{O}(\unit[10]{TeV})$ in the tau sector~\cite{Davidson:2022jai}. 

\section{Lepton flavor violation by two units}

The ``second ring'' around the origin in Fig.~\ref{fig:LFVgrid}, in which at least one lepton flavor is violated by \textit{two units}, has received less attention, despite the fact that these, too, are generated by $d=6$ SMEFT operators and thus not any more suppressed than e.g.~$\mu\to e\gamma$, even without imposing any symmetries. \textit{Half} of these processes are currently under investigation, namely muonium-antimuonium conversion $\bar{\mu}e\to \mu \bar{e}$~\cite{Bai:2022sxq} and the tau decays $\tau \to \mu\mu \bar{e}$ and $\tau \to e e \bar{\mu}$. The other half break tau number by two units and thus inevitably involve two tau leptons, e.g.~$\ell_{e,\mu} \ell_{e,\mu}\to \tau \tau$. The two taus make it kinematically impossible to looks for these processes in tau decays and thus require different search strategies. Aside from the obvious choice of same-sign lepton colliders~\cite{Dev:2023nha}, one can look at four-body $Z$ decays, $Z\to \tau\tau \bar{\ell}_{e,\mu}\bar{\ell}_{e,\mu}$, although current limits only reach $\Lambda \sim \mathcal{O}(\unit{GeV})$, inconsistent with the SMEFT setup. Renormalizable models with light new particles could give testable rates for such $Z$ decays though~\cite{Altmannshofer:2016brv,Altmannshofer:2022fvz}, and future $Z$-factories might eventually probe relevant $\Lambda$ scales~\cite{Heeck:2024uiz}. 

Other processes can be obtained by letting one of the two taus be off-shell, which then allows for tau decays such as $\tau \to ee \bar{\tau}^*\to ee \bar{\ell} \nu_\ell \bar{\nu}_\tau$, with $\ell \in \{e,\mu\}$. Such decays have been searched for in CLEO~\cite{CLEO:1995azm} and provide similar limits as $Z$ decays~\cite{Heeck:2024uiz}.

Interestingly, if we focus on the underlying $d=6$ SMEFT operators themselves, for example~$\bar{L}_\tau \gamma^\alpha L_e \,\bar{\ell}_\tau \gamma_\alpha \ell_e$, we can often get good limits from processes that are not strictly speaking \textit{charged} lepton flavor violating. As long as the $\Delta L_\tau = 2$ operator contains a left-handed tau $SU(2)_L$ doublet $L_\tau$, the $SU(2)_L$ expanded operator will automatically come with tau neutrino operators. The above operator therefore not only generates $ee\to \tau\tau$, which is hard to test, but also $\tau\to e \nu_e\bar{\nu}_\tau$, which involves neutrinos and is thus not CLFV, but is still a good signature. The electron energy spectrum in all cases matches the Michel spectrum, so only the total $\tau\to e +\text{neutrinos}$ rate is affected, but tests of lepton flavor universality~\cite{Belle-II:2024vvr} still provide constraints of order $\Lambda \sim \mathcal{O}(\unit{TeV})$, applicable to the SMEFT~\cite{Heeck:2024uiz}. Most  $\Delta L_\tau = 2$ CLFV operators are currently best constrained through these surprising deviations from lepton flavor universality! In one way or another, we hence have experimental limits on the first \textit{and second} ring of Fig.~\ref{fig:LFVgrid}.

\section{Lepton flavor violation by three units}

Let us boldly push further and look at the third ring of Fig.~\ref{fig:LFVgrid}, only partially shown, in which at least one lepton flavor is violated by \textit{three} units. Even though we can easily impose a symmetry that would make them the dominant CLFV process, e.g.~$U(1)_{L_e + 4 L_\mu - 5 L_\tau}$ to forbid everything but $\tau \to eee\bar{\mu}\bar{\mu}$, the underlying operators start at $d=10$ in the SMEFT and are hence far more suppressed by $\Lambda$ than the first- and second-ring processes discussed so far. At this point, it is far from clear that they could even be experimentally observable, despite tau decays such as $\tau \to eee\bar{\mu}\bar{\mu}$ being fairly clean signatures in $B$ factories, likely testable down to branching ratios of order $10^{-9}$.

As discussed in Ref.~\cite{Heeck:2025jfs}, testable rates for $\tau \to eee\bar{\mu}\bar{\mu}$ require light new particles to overcome phase space and $\Lambda$ suppression, for example through the decay chain 
$\tau \to ee\bar{\mu} S$, $S\to e \bar{\mu}$, where $S$ is an SM singlet scalar with mass between $m_\mu$ and $m_\tau$. The decay length of $S$ can be short enough for this decay chain to appear prompt, and hence mimic $\tau \to eee\bar{\mu}\bar{\mu}$. Despite many indirect constraints on this model, $\tau \to ee\bar{\mu} S$ with branching ratio or order $10^{-9}$ appear feasible for $m_S \sim m_\mu$, potentially making this $\Delta L_\mu = 3$ setup testable at Belle II. The light new particles required for such large rates also give rise to novel collider signatures that are interesting in their own right.

As a side note, operators that violate one lepton flavor by three units already arise at $d=7$ in the SMEFT, albeit together with total lepton number, e.g.~$\bar{\tau} L_e L_e L_e H$. Due to the flavor content, these $\Delta L = 2$ operators do \textit{not} induce Majorana neutrino masses. These operators all induce simple $\ell_\alpha\to \nu\nu \ell_\beta$ decays, which hide the lepton flavor and number violating details in the neutrinos. Unlike the CLFV-by-two-units examples from above, these three-body decays modify the Michel spectrum of the outgoing lepton and can thus be constraint through limits on Michel parameters, which reach $\Lambda \sim \unit[700]{GeV}$ for muons and $\Lambda \sim \unit[300]{GeV}$ for taus~\cite{Heeck:2025jfs}. Once again, we get good limits on lepton-flavor-violating SMEFT operators by looking at seemingly flavor-\textit{conserving} processes.

\section{Conclusion}

Lepton flavor violation in the tau sector is a fertile ground for new-physics searches. Not only is it necessary for us to observe tau flavor violation to understand how the SM flavor symmetry $U(1)_{L_\mu-L_\tau}\times U(1)_{L_\mu + L_\tau - 2 L_e}$ is broken, the large tau mass also makes it possible to probe a vast number of CLFV SMEFT operators in tau decays.
The majority of CLFV studies focus on flavor violation by \textit{one} unit, e.g.~$\ell_\alpha\to\ell_\beta \gamma$, even though these needn't be the dominant processes. CLFV by \textit{two} units arises at the same $d=6$ mass dimension in the SMEFT but requires different search strategies, especially for $\Delta L_\tau =2$ operators, which are the most difficult ones to test. Precision measurements of $\tau\to \ell \nu\nu$ turn out to provide crucial constraints. 
CLFV by \textit{three} units is more complicated still and requires light-new-physics models to induce testable rates for the rather clean channels $\tau \to eee \bar{\mu}\bar{\mu}$ and $\tau \to \mu\mu\mu \bar{e}\bar{e}$.

Until we have pinned down how lepton flavor is violated in nature, we have to search as far and wide as we can to ensure not missing new physics due to theoretical bias. Studies of tau decays provide unique opportunities for this endeavor and we strongly encourage our experimental colleagues to explore novel corners of model space.

\section*{Acknowledgements}

I thank Mikheil Sokhashvili and Anil Thapa for collaboration on some of the projects presented here, and the organizers of TAU2025 for an interesting workshop!
This work was supported in part by the U.S. Department of Energy under Grant No.~DE-SC0007974.

\newpage
\bibliographystyle{utcaps_mod}
\bibliography{references}

\end{document}